\documentstyle[twoside,fleqn,espcrc2]{article}
\input epsf
   \newcommand{\eq}{\begin{equation}}
   \newcommand{\en}{\end{equation}}

\newcommand{\AmS}{{\protect\the\textfont2
  A\kern-.1667em\lower.5ex\hbox{M}\kern-.125emS}}

\title{
 Bound states in the 3d Ising model and
 implications for QCD at finite temperature and density}

\author{M.Caselle\address{Dipartimento di Fisica Teorica dell'Universit\`a di 
Torino and INFN,
Torino, Italy}, 
M.Hasenbusch\address{NIC/DESY Zeuthen, Germany},
P.Provero\address{Dipartimento di Scienze e Tecnologie Avanzate,
Universit\`a del Piemonte Orientale, Alessandria, Italy},
K.Zarembo\address{Department of Theoretical Physics, Uppsala University,
Uppsala, Sweden}} 
       
\begin{document}

\begin{abstract}
  We study the spectrum of bound states of the three dimensional Ising 
  model in the $(h,\beta)$ plane near the critical point. 
  We show the existence  of an unbinding line,
  defined as the boundary of the region where bound states exist.
 Numerical evidence suggests that this line coincides with the
$\beta=\beta_c$ axis. When the $3D$ Ising model is considered as an
effective description of hot QCD at finite density, we
conjecture the correspondence between the unbinding line and the line
that separates the quark-gluon plasma phase from the superconducting
phase. The bound states of the Ising model are conjectured to
correspond to the diquarks of the latter phase of QCD.
 \end{abstract}

\maketitle

\section{Introduction}
The use of effective models to describe the relevant degrees of freedom 
has always been an important tool in 
the study of non perturbative physics of gauge theories.
For pure gauge theories with a second order deconfinement transition,
the problem of finding such an effective description is solved by the
Svetitsky-Yaffe conjecture ~\cite{sy}.
\par
The case of finite density QCD is not as well understood. 
It is by now clear that in the phase diagram  of high $T$
and finite density QCD a central role is played by the second order phase 
transition located at the end of the first order line which separates the
hadronic phase from the quark-gluon plasma one at high density (see fig.1).
It has been conjectured that this point 
is
in the same universality class
as the order-disorder transition of the 3d Ising model. This identification is
even more important than the Svetitsky-Yaffe one for the zero density
case, 
since the difficulty of directly simulating finite density QCD makes
an effective description in terms of Ising-type variables more
valuable.

The critical behaviour of the 3d Ising model is by now very well
understood both 
numerically and field-theoretically.
In particular we have a good understanding of the properties of the free energy
and its derivatives near the critical point.

However, in view of an application to finite
density QCD one is more interested in phenomena
(like the Kert\'esz singularity~\cite{kert} or the
unbinding
line which we shall discuss below) which are {\sl not} 
related to any singularity
in the free energy.

The 
goal of this contribution is 
to discuss these quantities,
their behaviour in the 3d Ising model and the possible implications for finite
density QCD.  
Here we just state our results: for all details and derivations, see
~\cite{xxx1,xxx2}.

\section{Bound states in the 3d Ising model}
The $3D$ Ising model in the broken symmetry phase has a rich
spectrum
of non perturbative states, which can be interpreted as
bound states of the fundamental massive excitation~\cite{xxx1}.

This spectrum is mapped by duality into the glueball spectrum 
of the gauge Ising model, and hence includes states with all the
values of angular momentum $J$ and
parity $P$ allowed by the lattice structure. However for our current
purposes we can focus on the simplest one, that is the $J^P=0^+$ bound 
state of two elementary quanta.

At vanishing magnetic field, numerical simulations of the Ising model
show the existence of a state with mass $\sim 1.83\, m$, where $m$ 
is the 
mass of the fundamental excitation (inverse of the correlation length).  
On the other hand,
the Bethe-Salpeter equation 
for $3D$ $\phi^4$ theory in the broken symmetry phase predicts the
existence of a bound state of two fundamental quanta, 
with binding energy given at leading
order  by
\eq
E_b\sim m\exp\left({-\frac{8\pi m}{g}}\right)\sim 0.17\, m
\en
The excellent agreement between these numbers allows us to interpret
the 
state at $\sim 1.83 m$ as a bound state.

The Bethe-Salpeter equation shows also that when an external magnetic
field is switched on the binding energy decreases. Moreover,
in the unbroken symmetry phase the interaction between fundamental
excitations is repulsive, and no bound states can exist.  
These two facts suggest the
 existence of an {\em unbinding line} in the $(T,H)$ plane
delimiting the region where bound states exist. On the line the
binding energy vanishes. High precision Monte Carlo data confirm the
existence of this line and strongly suggest that it coincides with the 
$T=T_c$ axis.

\section{``Critical'' and pseudocritical lines in the $(H,T)$ plane.}

The 
unbinding line
$E_b(H,T)=0$ shares many features 
with the
{\em Kert\'esz line}, 
along which 
the cluster surface tension 
$\Gamma(H,T)$ vanishes
(see~\cite{wang} for the precise definition). 
Given a scaling function $X(h,t)$, expressed in terms of the reduced
variables  $h\equiv\beta H$ and 
$t=\frac{T-T_c}{T_c}\equiv\frac{\beta_c-\beta}{\beta}$,
a simple renormalization group argument shows that the line along
which $X$ vanishes must have the form
\eq
t=a_X |h|^{\frac{y_t}{y_h}}
\label{shape}
\en
where $y_t$ and $y_h$ are the RG eigenvalues of the energy
and spin operators.
The constant $a_X$  depends on the choice of the observable $X$.

Since both $E_b$ and
$\Gamma$ are scaling functions, both the unbinding line and the
Kert\'esz line will be described by Eq.~\ref{shape}. Numerically we have 
for the unbinding line
\eq
a_{E_b}\sim 0
\en
while for the Kert\'esz line
\cite{wang}
\eq
a_{\Gamma}\sim 0.39
\en
\par
This indicates that the two lines are definitely different. Both can be
considered as ``critical lines'' since they divide the $(h,t)$ plane into
well defined phases which can be distinguished by an order parameter ($E_b$ or
$\Gamma$) in the two cases, which is different from zero in one phase and
vanishes in the other phase. However these are not phase transitions
in the usual sense since the free energy is not singular on either of
these lines.

It is interesting to compare the Kert\'esz and 
unbinding line 
to the so called
{\em pseudocritical} lines,
defined as the loci of the maxima in the $(H,T)$ plane of quantities, like the
susceptibility $\chi(h,t)$, the specific heat $C(h,t)$ 
or the correlation length $\xi(h,t)$ which diverge at the
critical point.
Since also these quantities are scaling functions, 
the functional form of the
pseudocritical lines must be the same of eq.(\ref{shape}),
with
various constants $a_{\chi}, a_{C}, a_\xi,...$.


\par

It is very interesting to compare the various values of these constants. It
seems that the pseudocritical lines divide into two families, which are, so to
speak, ``attracted'' by the Kert\'esz and 
unbinding
lines respectively.
In particular it turns out that $a_\chi\sim a_\xi \sim a_\Gamma$ and $a_C\sim
a_{E_b}$.  While the susceptibility pseudocritical 
line lies near the Kert\'esz line
but is definitely different from it (a similar behaviour was recently observed
also in d=2~\cite{fortunato}) the pseudocritical line related to the correlation
length seems to almost coincide with the Kert\'esz line. It would be very
interesting to understand if this is only a coincidence, if a higher resolution
analysis separates the two lines and if a similar phenomenon also holds in other
models or in the $2d$ Ising model.

\section{Implications for finite density QCD.}
Our current understanding of finite $\mu$ and $T$ QCD is well summarized by fig.
1 (taken from~\cite{hands}).
\begin{figure}[htb]
\centerline{
\setlength\epsfxsize{230pt}
\epsfbox{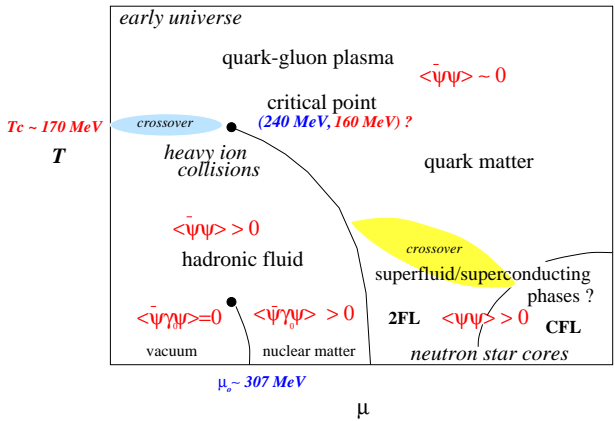}
}
\caption{Schematic view of the QCD phase diagram.}
\label{fig:qcd}
\end{figure}

In particular, as $\mu$ increases we recognize three regimes: 
(1) At low $\mu$, since the chiral symmetry is not exact the hadronic phase
is separated from the quark-gluon plasma (QGP) one by a smooth crossover.
(2) At intermediate values of $\mu$
the two phases are separated by a first
order line whose ending point belongs to the Ising universality class.
(3) At large $\mu$ a new ``superconducting'' (SC)
phase appears in which quarks
pair and can form a condensate.

Again the SC phase is separated by the QGP only 
by a smooth crossover since both
are characterized by the same global symmetries. The boundary between the two is
only defined by the fact that the binding energy between quarks, 
which is~\cite{son99}
\eq
E_b\sim \frac{b}{g^5}\exp\left( -{\frac{3\pi^2}{\sqrt2 g}}\right)
\en
becomes zero. 

If the endpoint of the line of first order transition is in the Ising
universality class, then all critical indices and universal amplitude
ratios coincide with the Ising ones. It is natural to wonder whether
this identification provides us with useful insight about the
physically interesting crossover phenomena around the critical point.
In this spirit, it was conjectured in \cite{fs} that the Kert\'esz line
is to be identified with the separation between hadronic and
QGP phase.
Here we propose to identify  the
unbinding line with the line separating  the SC and QGP phases.
More precisely, we suggest to identify the bound states of
elementary quanta of the Ising model with the diquarks of the SC phase. This
identification is supported by the similar behavior of the two binding
energies which both depend 
exponentially on the inverse coupling. Moreover both in the Ising and
in the QCD case the two phases have the same global symmetry and the only
order parameter which allows to
distinguish between them is the binding energy. 

Let us stress however that at this stage this identification (as the one which
involves the Kert\'esz line) is only a conjecture and is by no means 
implied by universality. Thus it would be very interesting
to study it in other 
models belonging to the Ising universality class,
which could better represent
the QCD phase diagram,
like the 3d Potts model in a magnetic field considered
in~\cite{ks}.

\end{document}